# New radical cation salt $\kappa$-(BETS)$_2$Co$_{0.13}$Mn$_{0.87}$[N(CN)$_2$]$_3$ with two magnetic metals: synthesis, structure, conductivity and magnetic peculiarities.

N.D. Kushch[1], O.M. Vyaselev[2*], V.N. Zverev[2,3], W. Biberacher[4], L.I. Buravov[1], E.B. Yagubskii[1], E. Herdtweck[5], E. Canadell[6], M.V. Kartsovnik[4]

[1]*Institute of Problems of Chemical Physics, Russian Academy of Sciences, 142432, Chernogolovka, Russia;*

[2]*Institute of Solid State Physics, Russian Academy of Sciences, 142432, Chernogolovka, Russia;*

[3]*Moscow Institute of Physics and Technology, 141700, Dolgoprudnyi, Russia;*

[4]*Walther-Meissner-Institut, Bayerische Akademie der Wissenschaften, Walther-Meissner-Strasse 8, Garching D-85748, Germany;*

[5]*Technische Universität München, 85747 Garching, Lichtenberg Str. 4, Germany;*

[6]*Institut de Ciència de Materials de Barcelona (CSIC), Campus de la U.A.B., E-08193 Bellaterra, Spain;*

**Abstract**

A new metallic radical cation salt $\kappa$-(BETS)$_2$Co$_{0.13}$Mn$_{0.87}$[N(CN)$_2$]$_3$, where BETS is bis(ethylenedithio)tetraselenafulvalene, C$_{10}$S$_4$Se$_4$H$_8$, has been synthesized. In this salt, a part of Mn$^{2+}$ ions are replaced by Co$^{2+}$ which acts as a magnetic dopant with a different effective magnetic moment. Crystal structure, band structure, conducting and magnetic properties of the salt have been studied. Below 30 K the material undergoes a metal-insulator transition, which is suppressed by applying a pressure of ~ 0.5 kbar, leading to a superconducting ground state. While the structural and conducting properties are very similar to those of the parent salt $\kappa$-(BETS)$_2$Mn[N(CN)$_2$]$_3$, magnetic properties associated with localized moments in the anion layer are found to be surprisingly different.

**1. Introduction**

Design and synthesis of new hybrid materials with pre-specified practical properties is one of the most interesting and promising areas of modern science. A combination of two (or more) externally controllable physical properties within the same compound allows expanding the material potentialities for the electronics of the future. In the past decade, large interest has been attracted to the design of hybrid organic compounds that combine conductive and magnetic properties [1-9]. From this viewpoint, the most promising compounds are the radical cation salts based on π-electron donors with paramagnetic metal complex anions. The conductivity in these compounds is provided by π-electrons of the organic radical cation layers and magnetic properties are associated with localized spins of the *d*- or *f*- metals incorporated in the anion layers. Cation radical salts of

---

[*] Corresponding author. e-mail: vyasel@issp.ac.ru; Fax: +7 49652 4 97 01.

bis(ethylenedithio)tetrathiafulvalene (BEDT-TTF) and its selenium-substituted analog BETS, are especially interesting. These salts were found to exhibit ferromagnetism [10, 11], as well as superconductivity coexisting with paramagnetic [12-14] and antiferromagnetic [15-17] properties. Furthermore, it was found that the interaction between localized spins in the anionic insulating layers with spins of π-electrons in the conducting layers may be accompanied by such an unusual phenomenon as magnetic field induced superconductivity. The latter was observed, for example, in λ-(BETS)$_2$FeCl$_4$ [18] and κ-(BETS)$_2$FeBr$_4$ [19]. The recently synthesized BETS dicyanamidomanganate with the paramagnetic $Mn^{2+}$ ion, κ-(BETS)$_2$Mn[N(CN)$_2$]$_3$ (hereafter abbreviated as κ-BETS-Mn), has attracted much interest. At ambient pressure the salt undergoes a metal-insulator transition at $T \sim 22$ K [20]. As the external pressure, $P$, is applied, the temperature of MI transition decreases and at $P > 0.5$ kbar a superconducting transition with $T_c = 5.8$ K is observed [20, 21]. It has been shown that π electron spins in the insulating state localize forming a long-range order of antiferromagnetic (AF) type, while the applied pressure restores the paramagnetic state of π-spins intrinsic to the metallic state of κ-BETS-Mn [22, 23].

As for the anion sublattice, below the MI transition temperature the $d$-electron $Mn^{2+}$ spins show a tendency towards AF ordering. However, no static long-range order as in λ-(BETS)$_2$FeCl$_4$ [18], occurs in the anion layer [23, 24], probably due to frustration in the triangular arrangement of $Mn^{2+}$ ions. Details of the atypical structure of $Mn^{2+}$ spins in the insulating state, neither purely paramagnetic nor AF, which is presumably induced by the AF-ordered π-spins through the π-d interaction [24, 25], is thus far unclear. There are, to the best of our knowledge, no reports related to synthesis of radical cation salts with other magnetic BETS-dicyanamidometallates which could help to clarify this issue. All our attempts to obtain crystals of radical cation salts κ-BETS$_2$M[N(CN)$_2$]$_x$ (M = Co, Cu, Fe, Ni, Cr; x = 3 or 4) in different solvents, using various techniques have failed: crystals either did not grow at all or had no magnetic metal in their composition.

In this work, we report on a successful synthesis of the new salt κ-(BETS)$_2$Co$_x$Mn$_{1-x}$[N(CN)$_2$]$_3$ with x = 0.13 (hereafter referred to as κ-BETS-CoMn) containing $Co^{2+}$ as a magnetic dopant. We present the crystal and band structures of this salt as well as its transport and magnetic properties and compare them to those of the κ-BETS-Mn salt.

2. **Experimental**

2.1. Synthesis

2.1.1. Electrolytes

The (Ph$_4$P)Mn[N(CN)$_2$]$_3$ and [(Ph$_4$)P]Co[N(CN)$_2$]$_3$ complexes used as electrolytes were prepared according to the procedure described in Ref. [26].

2.1.2. Crystals of κ-BETS-CoMn salt

The diamond shaped crystals of κ-BETS-CoMn were prepared in argon atmosphere by electrochemical oxidation of BETS (6.3 mg, 0.011 mmol) in a medium of benzonitrile – abs. ethanol. The reaction was performed at constant current $I = 0.5$ μA at 25°C using a mixture of electrolytes consisting of $[(Ph_4)P]Co[N(CN)_2]_3$ and $[(Ph_4)P]Mn[N(CN)_2]_3$ salts taken in equal amounts of 10 mg (0.017 mmol) each. The synthesis was carried out in a glass H-like two-electrode cell with cathodic and anodic chambers separated by a porous glass membrane. The electrodes were 1 mm-diameter platinum wires, electrochemically purified in a 0.1 N sulfuric acid solution.

Initially, the cell was flushed by argon through the porous partition from the anodic chamber to the cathodic one. Under the argon flow maintained, the cathodic chamber was filled with the solvents (20 ml of benzonitrile and 2 ml of abs. ethanol). 20 minutes later, as the solvents became saturated with argon, the electrolytes were added to the cathode chamber and the mixture was stirred for 40 minutes. Then the argon pressure in the cathodic and anodic chambers of the cell was equalized with a bypass valve to let the solution distribute between the chambers. Subsequently, BETS was added to the anodic chamber and stirred until the color of the solution turned light pink. The crystals grew on the anode for 2.5 weeks. Finally, the crystals were filtered off, washed with acetone and dried in air.

The synthetic procedure and the solvent amounts for BETS electrocryctallization in the presence of unequivalent amounts of the electrolytes was the same as described above. For 1:2 ratio we took 7 mg (0.012 mmol) of $[(Ph_4)P]Mn[N(CN)_2]_3$, 14 mg (0.023 mmol) of $[(Ph_4P)]Co[N(CN)_2]_3$ and 7 mg (0.012 mmol) of BETS; for 1:5 ratio — 3 mg (0.005 mmol) of $[(Ph_4)P]Mn[N(CN)_2]_3$, 15 mg (0.025 mmol) of $[(Ph_4P)]Co[N(CN)_2]_3$ and 6.5 mg (0.011 mmol) of BETS.

2.2. Electron-probe X-ray microanalysis (EPMA)

Preliminary composition of the salts was determined from the electron-probe X-ray microanalysis (EPMA) on a JEOL JSM-5800L scanning electron microscope (SEM) at 100-fold magnification and 20 keV electron beam density. The depth of beam penetration to the sample was 1-3 μm.

2.3. Single crystal X-ray analysis

X-ray analysis of the crystals was performed on a Bruker APEX-II CCD diffractometer (a tube with a rotating anode FR591, graphite monochromator, MoK$_α$ radiation, $λ = 0.71073$ Å) at 123 K. An array of experimental intensities from the crystals was refined to absorption by a multiple use of the SADABS program [27]. The structures were solved by direct method using SHELXS-90 [28] and refined by the full-matrix least-squares method using SHELX-97 [29]. Non-hydrogen atoms were refined in an anisotropic approximation. Parameters of hydrogen atoms were set geometrically using the "riding" model. Since the positions of Mn and Co atoms in the crystal lattice are the same, their proportion has been found from the refinement of their occupations, providing that their total occupation is 1. The main crystallographic data and parameters of structure refinement are presented in Table I.

### 2.4. Resistivity measurements

Sample resistance was measured using a four-probe technique and a lock-in amplifier at 20 Hz alternating current. The samples were thin plates with a characteristic lateral size of about 0.5 mm and the thickness in the range 20-50 μm. The surface of the plate was oriented along the conducting layers, i.e. parallel to the (*bc*) plane. Two contacts were attached to each of two opposite sample surfaces with conducting graphite paste. In the experiment we have measured the out-of-plane resistance $R_\perp$. The

| Table I. Crystallographic data and refined structural parameters for κ-BETS$_2$Co$_{0.13}$Mn$_{0.87}$[N(CN)$_2$]$_3$ ||
|---|---|
| Empirical formula | C$_{26}$H$_{16}$Co$_{0.13}$Mn$_{0.87}$N$_9$S$_8$Se$_8$ |
| Molecular weight | 1398.11 |
| $T$, K | 123(1) |
| Crystal size, mm | 0.46 x 0.36 x 0.05 |
| Crystal system | Monoclinic |
| Space group | P2(1)/c |
| $a$, Å | 19.4307(5) |
| $b$, Å | 8.3884(2) |
| $c$, Å | 11.9289 (2) |
| α, ° | 90 |
| β, ° | 92.416(1) |
| γ, ° | 90 |
| $V$, Å$^3$ | 1942.59(8) |
| Z | 2 |
| $\rho_{calc}$, g/cm$^3$ | 2.390 |
| μ, mm$^{-1}$ | 8.316 |
| F (000) | 1178 |
| Total number of reflections | 3581 |
| Number of independent reflections | 3113 |
| Number of observed reflections with $I \geq 2\sigma(I)$ | 3360 |
| Number of refined parameters | 252 |
| $R$-factor and wR factor for independent reflections | 0.0186 / 0.0548 |
| $R$-factor and wR factor for observed reflections | 0.0153 / 0.0388 |
| GOF | 1.038 |

measurements in the temperature range 1.3-300 K were carried out in a 4He cryostat with a variable temperature insert. To suppress the metal-insulator transition the samples of κ-BETS-CoMn were subjected to a quasi-hydrostatic pressure up to 3 kbar using the Cu-Be clamp cell with silicon oil as a pressure medium and with the manganin probe for the pressure control.

### 2.5. Electronic band structure calculations

The tight-binding band structure calculations were of the extended Hückel type [30]. A modified Wolfsberg-Helmholtz formula was used to calculate the non-diagonal H$_{\mu\nu}$ values [31]. All valence electrons were taken into account in the calculations and the basis set consisted of Slater-type orbitals of double-ξ quality for C 2s and 2p, S 3s and 3p, Se 4s and 4p and of single- ξ quality for H. The ionization potentials, contraction coefficients and exponents were taken from previous works [32, 33].

### 2.6. Magnetic properties

The dc magnetization of a 154 μg single crystal was measured using a Quantum Design MPMS XL-7 SQUID magnetometer, in 10 kOe field for the temperature range 20-300 K and in 1 kOe for $T$ = 2-30 K. The measurements were made for the field applied along the in-plane crystallographic directions *b* and *c*, and along *a\** perpendicular to (*bc*). The sample was fixed on a $4\times4\times0.2$ mm$^3$ Si substrate using Apiezon N grease. Magnetic moment of the substrate was measured separately and subtracted from the total measured magnetization.

Magnetic torque was measured on a 120 µg crystal in field sweeps up to 150 kOe using a homemade capacitive cantilever beam torque meter described in Ref. [34]. The cantilever was made of 25 µm thick as-rolled beryllium-copper foil. The torque meter was assembled on a single-axis rotatable platform in a way to pick up the torque component along the rotation axis directed perpendicular to the applied field. The device was calibrated by measuring the angular dependence of its capacitance in zero applied field, that is the torque caused by the gravity force, which was calculated using the known masses of the sample and the cantilever work platform.

## 3. Results and Discussion

3.1. Synthesis

As it was mentioned in the Introduction, salts based on BETS with cobalt dicyianamide anion have not been described in the literature. However, it is known that tetraphenylphosphonium dicyanamide salts of manganese and cobalt used in the synthesis of the radical cation salts as electrolytes are isostructural to each other. This implies a feasibility of κ-BETS-Mn dicyanamidometallate salts containing small cobalt additives. Mn$^{2+}$ in κ-BETS-Mn salt was determined to be in the high-spin state with $L = 0$, $J = S = 5/2$ [20, 22]. On the other hand, in octahedral complexes Co$^{2+}$ is usually in the high-spin ($S = 3/2$) state with $L \neq 0$ [35]. Therefore, Co-doped κ-BETS-Mn can be used to study the influence of a small addition of ions, having a different effective magnetic moment, on the anion magnetic subsystem properties as well as the overall effect of such doping on the system. Besides, it appears interesting to estimate the effect of chemical compression of the Co-doped salt, κ-BETS-CoMn, since Co$^{2+}$ ionic radius is smaller than Mn$^{2+}$.

All syntheses of the κ-BETS-CoMn salt were carried out in a medium of benzonitrile – abs. ethanol at constant current (0.3-0.75 µA) in the electrolyte mixtures of [(Ph$_4$P)]Mn[N(CN)$_2$]$_3$ and [(Ph$_4$)P]Co[N(CN)$_2$]$_3$ taken at ratios from 1:1 to 1:5. The double excess of [(Ph$_4$)P]Co[N(CN)$_2$]$_3$ in the mixture yielded the BETS radical cation salt crystals with simple dicyanamide anions growing on the anode [36]. At the five-fold excess of [(Ph$_4$)P]Co[N(CN)$_2$]$_3$ the crystals did not grow at all. The use of equal amounts of Mn- and Co- tetraphenylphosphonium dicyanamide salts resulted in the desired cation-radical κ-BETS-CoMn salt. The composition of the salt was preliminarily determined from EPMA and finally refined by the full X-ray analysis. The distribution of the metals in the grown crystals was found by analyzing the EPMA data. The analysis showed that Co ions in the crystals are distributed inhomogeneously, while the average ratio of Mn and Co contents is close to 9:1.

## 3.2. Crystal structure

Crystal structure of the obtained κ-BETS-CoMn salt is illustrated in Fig. 1. Quite expectedly, it is isostructural to the parent manganese salt κ-BETS-Mn [20, 21].

κ-BETS-CoMn crystals have a layered structure characterized by the alternation of conducting radical cation- and insulating magnetic anion layers along *a* direction. The conducting layers consisting of BETS dimers belong to the κ-type packing. The dihedral angle between the dimers is 73.28° at 123 K. The corresponding angle in κ-BETS-Mn is 72.8° at 273 K [20], 73.10° at 110 K and 73.86° at 15 K [21]. It is clear that in κ-BETS-Mn the angle between the dimers increases with decreasing temperature that indicates some redistribution of the interactions between BETS molecules inside and outside the dimers. Apparently, a similar redistribution is also realized in κ-BETS-CoMn since the dihedral angle between the dimers here is already bigger at 123 K than in κ-BETS-Mn at 110 K. Comparison of the crystallographic cell parameters of κ-BETS-CoMn and κ-BETS-Mn [21] salts shows that *a* and *b* are practically identical, while *c* is only ~1.5% longer in κ-BETS-CoMn. This infers the absence of a noticeable chemical compression effect associated with the difference in ionic radii of $Co^{2+}$ and $Mn^{2+}$.

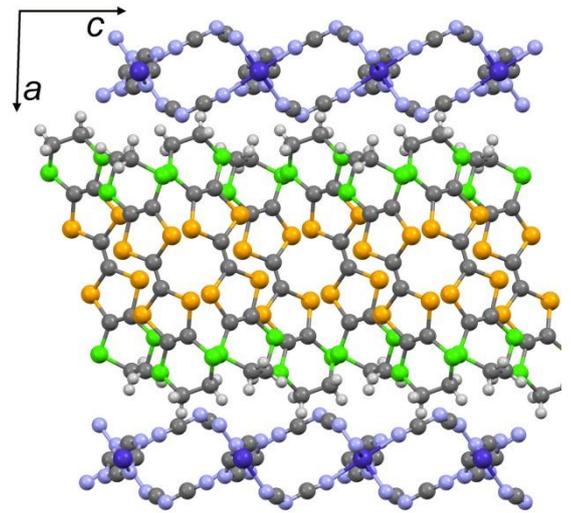

Figure 1 (Color online). Crystal structure of κ-BETS-CoMn projected on the *ac*-plane

It was shown [21] that the κ-BETS-Mn salt synthesized according to the procedure described in Ref. [20] undergoes a phase transition associated with formation of an incommensurate superlattice along *b* direction (wave vector $q = 0.42b$) in the vicinity of $T = 102$ K. This assumption is supported by a strong increase in the intensity of the superstructure reflections when cooling below the phase transition temperature. The superstructure remains unchanged down to 15 K. However, crystals of both κ-BETS-Mn and κ-BETS-CoMn salts grown according to another synthetic procedure [36] demonstrate no such one-dimensional modulated structure. This affects the normal-state shape of *R*(*T*) curves for both salts (for κ-BETS-CoMn see Section 3.4 below), but not the parameters of the superconducting transition. There are several short contacts S...Se, S...S and one hydrogen bond CH...S in the conducting layers of κ-BETS-CoMn. Besides, there are shortened contacts N...S and N...C as well as hydrogen bonds CH...N between BETS molecules and dicyanamide anion groups. The short intermolecular contact values in the structure of κ-BETS-CoMn salt are listed in Table II. The hydrogen bonds CH...N link the cation radicals BETS and the $Me[N(CN)_2]^{3-}$ (Me = Mn, Co) octahedrons that form a 2D polymeric anionic layers shown in Fig. 2. One of the two crystallographically independent dicyanamide groups is disordered by the amide nitrogen atom and both carbon atoms of the terminal groups C-N. It should be noted that in κ-BETS-CoMn there are fewer short contacts between the heteroatoms in the radical cation layers than in κ-BETS-Mn at any temperature. In contrast, the number of short intermolecular contacts between the conductive and insulating layers is larger in κ-BETS-CoMn. Note also that in κ-BETS-CoMn there are primarily the CH...N hydrogen bonds linking the conductive and magnetic layers (see Table II), while only CH...S type hydrogen bonds in the conducting layers are present in κ-BETS-Mn. Thus, the analysis of short intermolecular contacts in both salts indicates a stronger interaction between the radical cation- and anion layers in κ-BETS-CoMn as compared to κ-BETS-Mn.

Table II. Short intermolecular contacts in the structure of κ-BETS$_2$Co$_{0.13}$Mn$_{0.87}$[N(CN)$_2$]$_3$.

| Contact | Contact length, Å | Symmetry operation for the 2$^{nd}$ atom in contact |
|---|---|---|
| Se2…S5 | 3.511 | x, 0.5−y, 0.5+z |
| Se2…S8 | 3.585 | 1−x, 1−y, 1−z |
| Se3…S8 | 3.675 | x, 0.5−y, −0.5+z |
| S5…S6 | 3.579 | x, 0.5−y, −0.5+z |
| S6…S8 | 3.509 | 1−x, 1−y, 1−z |
| S7…S8 | 3.380 | 1−x, 0.5−y, −0.5+z |
| N1…S7 | 3.204 | 1−x, 0.5+y, 0.5−z |
| C7…N3 | 3.224 | 1+x, y, z |
| H7A…S7 | 2.792 | 1−x, −0.5+y, 0.5−z |
| H10B…N1 | 2.727 | 1+x, 1.5−y, 0.5+z |
| H8B…N2 | 2.732 | 1+x, 0.5−y, 0.5+z |
| H7B…N3 | 2.513 | 1+x, y, z |
| H7B…N4 | 2.522 | x, y, z |
| H9B…N5 | 2.665 | 1+x, 0.5−y, 0.5+z |

It is also possible that the presence or absence of the superstructure in the salts is determined by the synthesis conditions (especially the nature of the solvent and the electrolyte) that affects also the intermolecular contacts in the crystal, conformation of the donor molecule, the degree of the structural order etc., as it occurs, for example, in β-(BEDT-TTF)I$_3$ crystals [37] and in polymorphous series of the (BETS)$_2$Cu[N(CN)$_2$]Cl crystals [36,38,39]. It is known, for example, that if CuCl$_2$ is used instead of CuCl as an electrolyte component in the synthesis, under the same conditions a polymorph κ'-(BEDT)$_2$Cu[N(CN)$_2$]Cl grows on the anode instead of the radical cation salt κ-(BEDT)$_2$Cu[N(CN)$_2$]Cl [36].

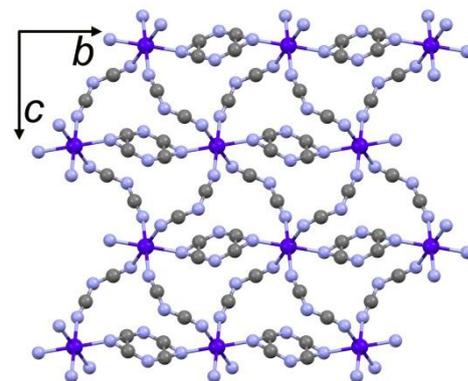

Figure 2 (Color online). Projection of the anion layer on the *bc*-plane

3.3. Electronic band structure

The donor layers of BETS contain only one symmetry inequivalent BETS donor molecule and four different types of donor..donor interactions (see Fig. 3 for the labeling). The calculated |β$_{HOMO-HOMO}$| values [39] for every type of intermolecular interaction, which are a useful measure of the strength of the interaction between the HOMOs of two adjacent donors, as well as the density of states at the Fermi level, $n(\varepsilon_F)$, calculated on the basis of the 123 K crystal structure of this salt, are given in Table III.

As it was observed [21] for the isostructural system κ-BETS-Mn the calculated |β$_{HOMO-HOMO}$| values clearly indicate that, from the electronic structure viewpoint, the donor layers of this salt are

more adequately described as a series of interacting chains of dimers along the b direction than as a genuine two-dimensional lattice of interacting donors, as it would be the case for most κ–phases. Note that the strength of the different interactions as well as the density of states at the Fermi level of the present salt at 123 K are in between those calculated for κ-BETS-Mn at 200 and 110 K [21]. Consequently, the donor layers of the present salt are very similar to those of the parent κ-BETS-Mn system. The reasons for the lesser two-dimensional character of the electronic structure were already examined for κ-BETS-Mn [21].

The calculated band structure and Fermi surface are shown in Figs. 4 and 5, respectively. Fig. 4 is the expected band structure for a κ-type phase with quite strong dimers. The Fermi surface shape is typical of the κ phases: it can be seen as a superposition of closed pseudo-ellipses with an area of 100% of the cross-section of the Brillouin zone. Because of the symmetry of the lattice there is no gap at the intersection of the two lines along the Z→M line. The area of the closed portion around Z is 27% of the cross-section. Again, this Fermi surface is very similar in shape to that of the parent κ-BETS-Mn salt.

The Fermi surface in Fig. 5 is somewhat atypical of those in κ-phases in that there are quite sizeable portions which are very flat. For instance, the closed portion around Z is almost a square. In addition, it is quite clear that a tiny change (lowering) of the Fermi level would change the shape near the Γ point so that there would be no more closed portions around Z. The Fermi surface would consist in that case of two series of warped lines touching at one point of the Z→M boundary of the Brillouin zone. These features indicate that, as it was the case for κ-BETS-Mn, the salt is more one-dimensional than the typical κ-phases. In the usual κ-phases the interaction between chains is stronger, leading to the typical two-dimensional behavior.

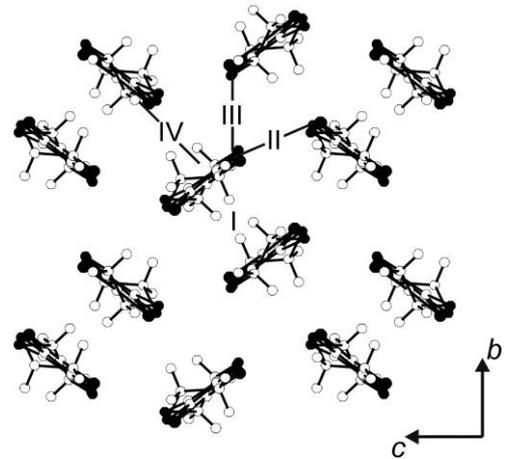

Figure 3 Donor layer of κ-BETS-CoMn where the four different types of donor..donor interactions are labeled.

Table III. Calculated values of the $|\beta_{HOMO-HOMO}|$ [eV] for the different donor..donor interactions and density of states at the Fermi level, $n(\varepsilon_F)$ [electrons/eV·unit cell], for κ-BETS-CoMn at 123 K. The values calculated for the isostructural κ-BETS-Mn salt at 200 and 110 K [21] are also given.

|  | κ-BETS-CoMn | κ-BETS-Mn | |
| --- | --- | --- | --- |
|  | 123 K | 200 K | 110 K |
| I | 0.6077 | 0.5883 | 0.6185 |
| II | 0.1221 | 0.1293 | 0.1196 |
| III | 0.3407 | 0.3365 | 0.3407 |
| IV | 0.0434 | 0.0373 | 0.0472 |
| $n(\varepsilon_F)$ | 8.47 | 8.50 | 8.41 |

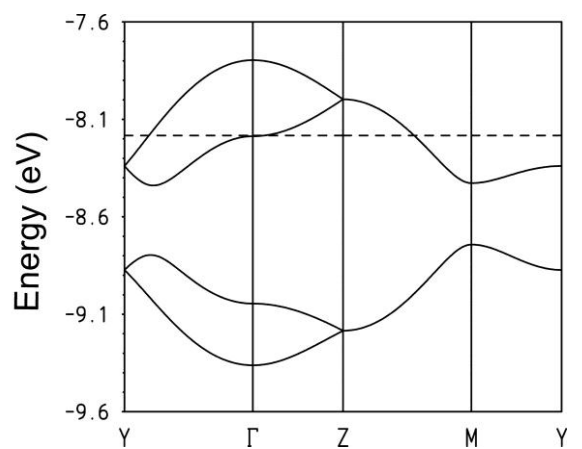

Figure 4. Calculated band structure for the donor layers of κ-BETS-CoMn at 123 K where the dashed line refers to the Fermi level and Γ = (0, 0), Y = ($b*/2$, 0), Z = (0, $c*/2$) and M = ($b*/2$, $c*/2$).

The Fermi surface in Fig. 5 exhibits nesting properties allowing the superposition of the opposite edges of the pseudo squares around Z with the nesting vectors $\mathbf{q} \approx 0.48\ \mathbf{c^*} \pm 0.28\ \mathbf{b^*}$. However, this will only suppress a part of the Fermi surface and, because of the possible changes of the Fermi surface shape around the Γ point, it is not clear if these pseudo squares will survive at low temperatures. The strong similarity in the electronic structures of κ-BETS-CoMn and κ-BETS-Mn supports the idea that the incommensurate modulation in the latter salt must originate either from the anion layer or in the way the donor and anion layers interact through hydrogen bonding.

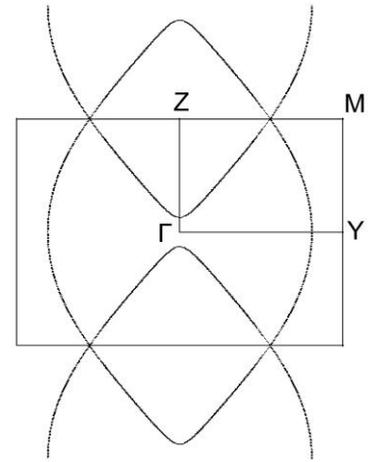

Figure 5. Calculated Fermi surface for the donor layers of κ-BETS-CoMn at 123 K.

Shown in Fig. 6, is the calculated density of states for the donor layers. As for κ-BETS-Mn, the Fermi level lies just at a peak of the density of states, something that is most likely related to the observation of superconductivity in the present phase once the Mott type metal-insulator [21] is suppressed.

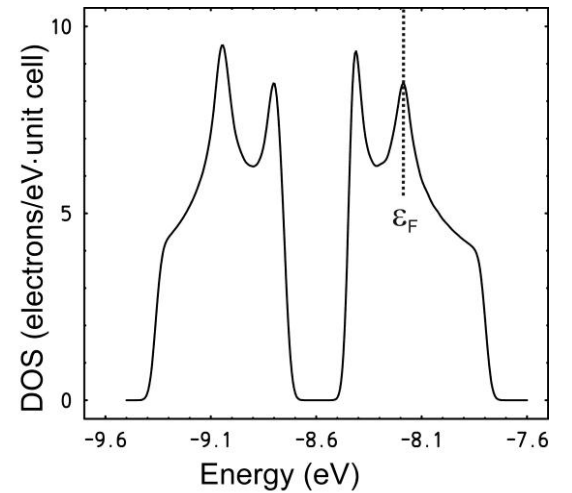

Figure 6. Calculated density of states (in units of electrons/eV·unit cell) for the donor layers of κ-BETS-CoMn at 123 K.

3.4. Conducting properties

The ambient-pressure interlayer resistivity $\rho_\perp$ was found to be ~ 10 Ohm·cm at room temperature and showed the same temperature dependence $R(T)$ for all samples tested. Examples of $R(T)$ curves at different pressures are given in Fig. 7. Hereafter we present the pressure values $P$ determined for temperatures below 20 K. No evidence of the superstructure transition, like those reported earlier for the κ-BETS-Mn salt [21], has been found at any pressure. The kink around ≈ 130 K on the $R(T)$ curve for $P$ = 0.5 kbar is caused by freezing of the silicon oil used as a pressure medium [40].

As seen in Fig. 7, the compound undergoes a metal-insulator transition below 30 K at ambient pressure. The transition is suppressed under the applied external pressure, and at about 0.5 kbar the sample becomes superconducting with the critical temperature $T_c$ (0.5kbar) = 4.6 K. The pressure dependence of $T_c$ is presented in Fig. 8; the value of $T_c$ was determined as the onset transition point on $R(T)$ curve. The superconducting transition temperature goes down as the pressure increases. For a comparison, the *T-P* diagram for the superconducting transition in κ-BETS-Mn from Ref. [21] is also presented in Fig. 8. At pressures below 1.5 kbar the critical temperature is ≈ 1 K lower for κ-BETS-CoMn. Taking into account that superconductivity is often sensitive to the crystal quality in the organics [41, 42], one can speculate that $T_c$ is suppressed because of enhanced scattering in the mixed salt, hinting at a possible nodal pairing symmetry. On the other hand, superconductivity seems to be more robust with respect to applied pressure: a weak superconducting drop can be traced at least up to ≈ 3 kbar.

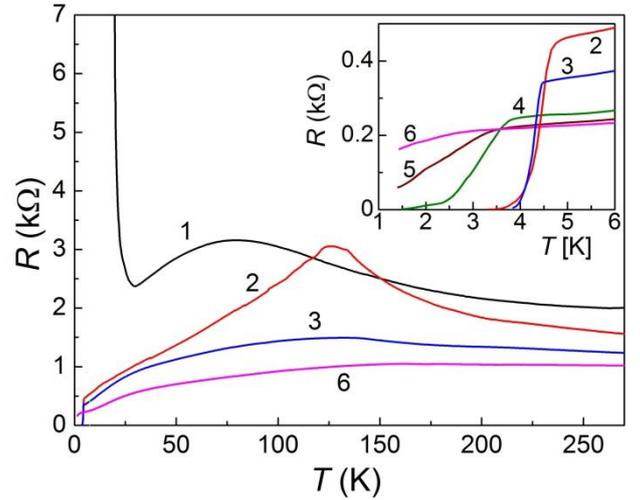

Figure 7 (Color online). Temperature dependencies of the interlayer resistance of κ-BETS-CoMn at different pressures, *P*. The inset demonstrates the pressure evolution of the superconducting transition. Curves 1 to 6: *P* = 0, 0.5, 0.9, 1.7, 2.2 and 2.9 kbar, respectively.

3.5 Magnetic Properties

The dc susceptibility of the sample determined as χ = *M*/*H* where *M* is the measured magnetization of the sample per mole, is nearly isotropic with χ = 0.45, 0.46 and 0.49 cm³/mol at *T* = 2 K for *H* along *a**, *b* and *c* directions, respectively. Fig. 9 shows the temperature dependence of χ along the *b* axis. The behavior of χ(*T*) is the Curie-Weiss – like, χ ∝ (*T*–$T_{CW}$)$^{-1}$, as evidenced by the straight-line χ$^{-1}$(*T*) plot in the top-left inset of Fig. 9. It was shown previously [20, 22] that the magnetization of the parent κ-BETS-Mn salt, is dominated by Mn$^{2+}$ in the high-spin state $t_{2g}^3 e_g^2$ (*L* = 0, *S* = 5/2). Assuming that in κ-BETS-CoMn the spin state of Mn$^{2+}$ is the same, we fit the susceptibility with the relation

$$\chi_{\text{calc}} = \frac{N_A \mu_B^2}{3k_B} \left( \frac{0.87 p_{\text{Mn}}^2}{T - T_{\text{CW}}^{\text{Mn}}} + \frac{0.13 p_{\text{Co}}^2}{T - T_{\text{CW}}^{\text{Co}}} \right), \qquad (1)$$

where $N_A$ is Avogadro's number, $\mu_B$ is the Bohr magneton, $k_B$ is the Boltzmann constant, $p_{\text{Mn}}$ and $p_{\text{Co}}$ are the effective numbers of $\mu_B$ for Mn and Co, respectively. For Mn$^{2+}$ we take $p_{\text{Mn}} = g\sqrt{S(S+1)} = 5.92$ with *S* = 5/2 and the Landé factor *g* = 2. The fit to the experimental data using $T_{CW}^{Mn}$, $T_{CW}^{Co}$, and $p_{Co}$ as fitting parameters yields nearly equal Curie-Weiss temperatures for Mn and Co, $T_{CW}^{Mn} = -5.8\,\text{K}$ and $T_{CW}^{Co} = -5.6\,\text{K}$, indicating the AF interactions. These values are practically identical to $T_{CW}^{Mn} = -5.2 \pm 0.7\,\text{K}$ obtained for κ-BETS-Mn [22]. The fact that $T_{CW}^{Mn} \approx T_{CW}^{Co}$ is an argument that Co does not form separate clusters with different magnetic properties. Parameter $p_{Co}$ obtained from the fits to the data is 4.1±0.3, 4.3±0.6 and 4.5±0.4 for $H\|a, b$ and $c$, respectively. In any case, $p_{Co}$ is no less then $\sqrt{15} = 3.87$ corresponding to $g = 2$, $S = 3/2$ which denotes the high-spin state of $Co^{2+}$. The orbital contribution from $Co^{2+}$ cannot be unambiguously resolved due to poor accuracy in determining $p_{Co}$ as Co contributes less than 10% to the total magnetic moment of the sample, which itself is rather small and spans from ~0.2×10⁻⁸ emu/Oe at 300 K to ~5×10⁻⁸ emu/Oe at 2 K. Eq. (1) fits the experimental data quite well, as shown by the solid line in the main panel of Fig. 9 and by the difference between the measured and calculated susceptibilities, $\chi-\chi_{calc}$, depicted on the top-right inset in Fig. 9. On the latter plot, the drop of $\chi-\chi_{calc}$ below $T \approx 7$ K is associated with diminishing of the measured $\chi$ in the vicinity of $T = -T_{CW}$, indicating a trend to the AF order. However, the observed deviation of $\chi$ from the Curie-Weiss behavior, which amounts ~10% at $T = 2$ K, is apparently insufficient to infer the long-range AF order of the Mn-Co spin subsystem. The $T$-dependencies of $\chi$ measured for the magnetic field applied along $a^*$ and $c$ directions are essentially the same as for $H\|b$ shown in Fig. 9, including the parameters $T_{CW}^{Mn} \approx T_{CW}^{Co} = -5.8 \pm 0.2\,\text{K}$ and $p_{Co} = 3.8 \pm 0.2$, as well as the characteristic diminishing of $\chi-\chi_{calc}$ below 7 K.

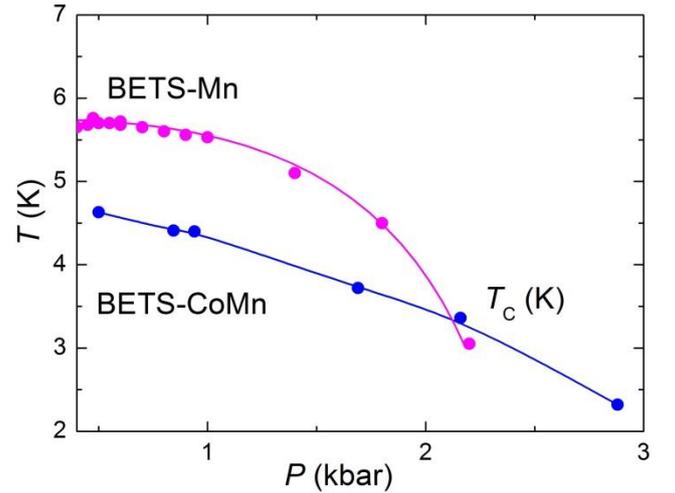

Figure 8 (Color online). Pressure dependence of the superconducting critical temperature for κ-BETS-CoMn and κ-BETS-Mn (from Ref. [21]). Lines are guides to the eye.

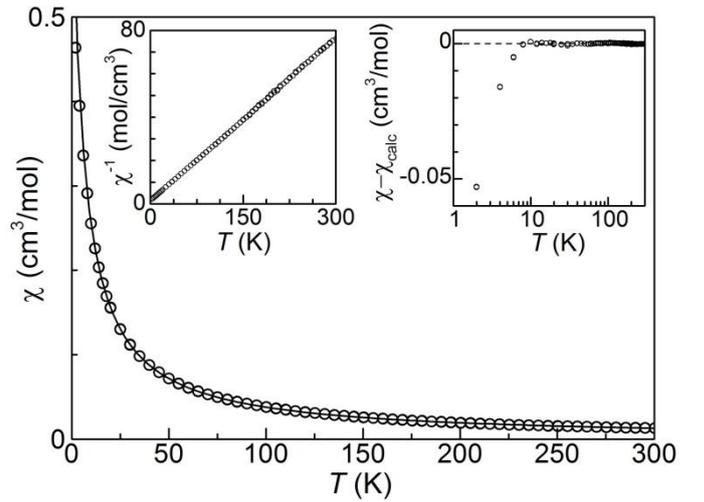

Figure 9. Main panel: Temperature dependence of the susceptibility of κ-BETS-CoMn for the magnetic field $H$ = 1 kOe ($T$ = 2–30 K) and 10 kOe ($T$ = 20–300 K), applied along $b$ direction. Open circles: experimental data; solid line: fit to the data, $\chi_{calc}$, using Eq. (1). Left inset: reciprocal susceptibility $vs.$ temperature. Right inset: difference between the measured and calculated ($\chi_{calc}$) susceptibilities.

The magnetic torque measured at temperatures above the metal-insulator transition, has a usual $H$-dependence of a paramagnet with a uniaxial anisotropy, $\tau \propto H^2\sin2(\theta-\theta_0)$, where $\theta$ is the magnetic field polar angle reckoned from $a^*$ direction. For both components of the torque in the ($bc$) plane, $\tau_b$ and $\tau_c$, we find, within experimental accuracy, $\theta_0 = 0°$. Thus, taking into account the crystal symmetry, we conclude that the magnetization principal axes of Mn-Co spin subsystem run along $a^*$, $b$ and $c$ directions, respectively. This differs from the parent κ-BETS-Mn compound [22, 25] where $\tau_b \propto \sin2(\theta-21°)$, indicating that the principal axes of Mn spin subsystem in the ($a*c$) plane are rotated by $\theta_0 = 21°$ from $a*$ and $c$.

The behavior of $\tau(H)$ curves becomes rather complicated as the system enters the insulating state. The magnetic torque components $\tau_b$ and $\tau_c$ measured at $T = 1.5$ K, are plotted against the applied field in Fig. 10. In both cases, $\tau(H)$ curves are non-monotonic and do not show any sign of saturation up to $H = 150$ kOe, unlike κ-BETS-Mn [22, 25] where $\tau(H)$ becomes constant at $H > 100$ kOe.

$\tau_b(H)$ curves [Fig. 10(a)] taken at different polar angles, θ, of the applied field cross each other at different field values, which signifies a complicate field evolution of the magnetic anisotropy. Fig. 11(a) shows the θ-dependencies of $\tau_b$ for $H = 46$, 85 and 150 kOe. It is obvious that none of the curves has a simple sin2θ-type pattern. It turns out, however, that $\tau_b(\theta)$ can be reasonably well described by adding a sin4θ component,

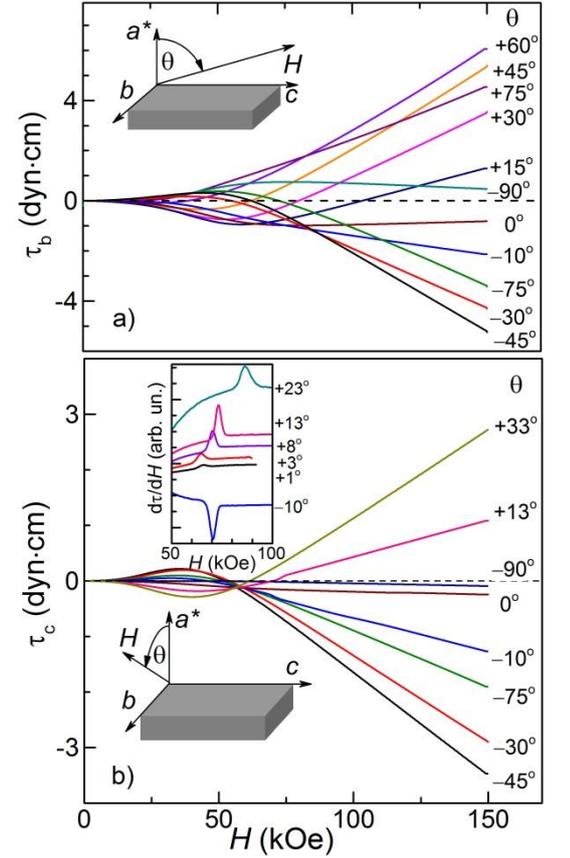

Figure 10 (Color online). Field dependencies of the magnetic torque components along (a) $b$ and (b) $c$ crystallographic axes, measured for different polar angles θ, at $T = 1.5$ K. Inset in (b): field derivatives of $\tau_c$ illustrating the presence of the spin-flop 'kinks' on $\tau_c(H)$.

$$\tau_b(\theta) = A\,[n \sin 2\,(\theta - \theta_0) + (n - 1) \sin 4\,(\theta - \theta_1)], \qquad (2)$$

as shown by solid lines in Fig. 11(a). Best fits to the data give $n = 0.75$, 0.76 and 0.83, $\theta_0 = 60°$, 18° and 4° for $H = 46$, 85 and 150 kOe, respectively. The phase of the 4θ harmonic, $\theta_1 = 45°$ is the same for all the three field values. One can see that as the field is increased, the θ-dependence of the torque approaches to that of the paramagnetic metallic state where $\theta_0 = 0$, $n = 1$. However, even at $H = 150$ kOe it is not completely restored, in contrast to κ-BETS-Mn where the $\tau(\theta)$ dependence at $T = 1.5$ K becomes identical to that of the metallic state already at $H = 70$ kOe.

The behavior of $\tau_c(H)$ curves [Fig.10(b)] looks more regular. At $H \approx 57$ kOe the $\tau_c(H)$ curves cross zero, which denotes the change of the sign of magnetic anisotropy. Figure 11(b) presents the angular dependencies of $\tau_c$ at 35, 57, and 150 kOe. As shown by solid lines, the data can be fitted by the conventional formula, $\tau(\theta) \propto \sin2(\theta-\theta_0°)$ at 35 kOe ($\theta_0 = 88°$) and 150 kOe ($\theta_0 = 0°$) fields, while at 57 kOe the torque is close to zero and nearly independent of θ. Thus, at high field the magnetic anisotropy is restored to its paramagnetic metallic state behavior, like in the parent κ-BETS-Mn salt.

The $\tau_c(H)$ curves recorded in the insulating state of κ-BETS-CoMn display step-like features ('kinks') at the field polar angles $|\theta| < 25°$, similarly to κ-BETS-Mn [22]. The kinks are better visualized on the field derivatives of $\tau_c(H)$, $d\tau_c/dH$, shown in the inset of Fig.10(b). They disappear upon warming up to the metallic state and are attributed to a field-induced spin-reorientation transition that occurs in the AF-ordered π-spin system in the insulating state [22]. The applied field values where the kinks occur for certain θ, are nearly the same as in κ-BETS-Mn, suggesting that the π-spin system properties in the two compounds are essentially identical.

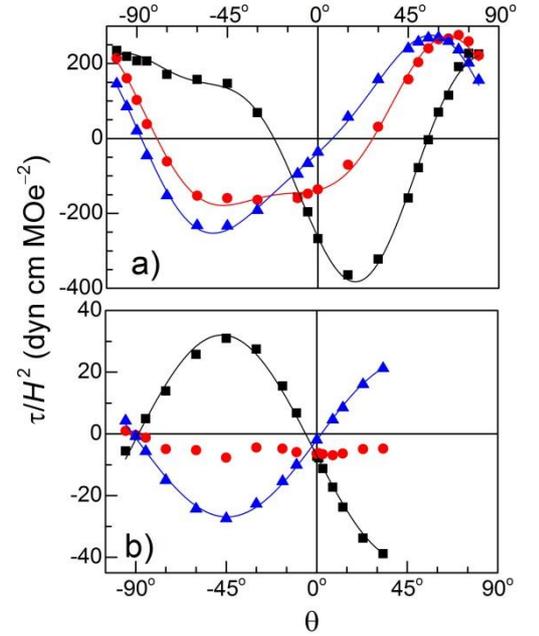

Figure 11 (Color online). Angular dependencies of the reduced torque, $\tau/H^2$, in κ-BETS-CoMn at $T = 1.5$ K. (a): $\tau_b(\theta)/H^2$ at $H = 46$ (squares), 85 (circles) and 150 kOe (triangles), solid lines are best fits to the data according to Eq. (2). (b): $\tau_c(\theta)/H^2$ at $H = 35$ (squares), 57 (circles) and 150 kOe (triangles), solid lines are best fits to the data using the relation: $\tau(\theta) \propto \sin2(\theta-\theta_0°)$.

Comparing the magnetic anisotropies in κ–BETS-CoMn and κ-BETS-Mn, we see considerable differences, both in the metallic and insulating states. In the metallic state the principal axes of magnetization in κ–BETS-CoMn coincide with $a^*$, $b$ and $c$ crystallographic directions, while in the parent salt they are rotated in the $(a^*c)$ plane by $\theta_0 = 21°$ from $a^*$ and $c$ [22, 25]. In both salts the anisotropy alters in the insulating state. In the parent salt the anisotropy retains the simple $\sin2\theta$ behavior; however the principal axes of magnetization are tilted by a few degrees [22, 25]. Under the applied field of ~ 70 kOe the tilt is removed and the magnetization principal axes return to their metallic-state directions. In the present salt the anisotropy changes more drastically in the insulating state: the angle dependence of the $b$-axis component of the torque, $\tau_b(\theta)$, acquires a complex shape [Fig. 11(a)], whereas the $\tau_c(\theta) \propto \sin2(\theta-\theta_0)$ dependence inverts at going from low to high fields [Fig. 11(b)]. Moreover, as mentioned above the anomalous behavior is much more robust with respect to the external field than in the parent salt. The reasons for such striking differences in the magnetic properties of the insulating states of otherwise very similar compounds are not clear at present.

## 4. Summary

In summary, crystals of a new radical cation salt κ-$(BETS)_2Co_{0.13}Mn_{0.87}[N(CN)_2]_3$ (κ–BETS-CoMn ) containing magnetic ions $Co^{2+}$ and $Mn^{2+}$ were synthesized. Its crystal and band structures are similar to those of the parent κ-BETS-Mn salt. In particular, like in κ-BETS-Mn, a large part of the Fermi surface is predicted to be almost flat and thus the conducting band should be more one-dimensional than in other κ-type salts of BETS and BEDT-TTF. At ambient pressure the κ–BETS-CoMn compound undergoes a transition into an insulating state with antiferromagnetically ordered π-electron spins at $T_{MI} \approx 21$ K. The transition is suppressed under an external pressure of 0.5 kbar, giving way to superconductivity with $T_c \cong 4.6$ K.

The dc magnetization of the system is dominated by $Mn^{2+}$ and $Co^{2+}$ ions, both in the high-spin state, and is well described by the Curie-Weiss law. The isotropic part of the magnetization is therefore determined by the same mechanism as in the parent κ-BETS-Mn salt. Surprisingly, the magnetic anisotropy shows a considerably different behavior from that in κ-BETS-Mn upon entering the insulating state. At present we do not have a reasonable explanation to the dramatic changes caused just by a 13% substitution of Co for Mn. On the other hand, properties of the π-electron system in the studied compound, including the temperature of the metal-insulator transition, AF ordering of π-spins in the insulating state and spin-reorientation transition fields, are very much the same as in κ-BETS-Mn. This suggests that spin systems of π-electrons and the paramagnetic ions of the anion layer are only weakly coupled.

**Acknowledgment**.

We thank Prof. A. Kobayashi for providing BETS used in the work. N.D.K. and E.B.Y. were supported by the RFBR grant No. 14-0300119 and by Program No. 2 of the Presidium of the Russian Academy of Sciences. N.D.K., O.M.V, W.B., and M.V.K. acknowledge support by the German Research Foundation (DFG) via the grant KA 1652/4-1. E.C. acknowledges support by MINECO (Spain) through Grant FIS2015-64886-C5-4-P, Generalitat de Catalunya (2014SGR301), and by the Spanish MINECO through the Severo Ochoa Centers of Excellence Program under Grant SEV-2015-0496.

**Supplementary material**

CCDC 1516720 contains the supplementary crystallographic data for this paper. These data can be obtained free of charge from the Cambridge Crystallographic Data Centre, http://www.ccdc.cam.ac.uk/data_request/cif.